\documentclass[aps,prb,preprint, amsmath,amssymp,showpacs]{revtex4-1}

\usepackage[T1]{fontenc}
\usepackage[latin1]{inputenc}
\usepackage[dvips]{graphicx,color}
\usepackage{rotating}
\usepackage{bm}        
\usepackage{amssymb}   
\usepackage{amsmath}
\usepackage{bm}
\usepackage{CJK}
\def\jcp#1#2#3{J.~Chem.~Phys.~{\bf #1},\ #2\ (#3)}

\def\pra#1#2#3{Phys.~Rev.~A~{\bf #1},\ #2\ (#3)}
\def\prl#1#2#3{Phys.~Rev.~Lett.~{\bf #1},\ #2\ (#3)}

\def\k1{k_1}
\def\k2{k_2}
\def\q1{q_1}
\def\q2{q_2}

\def\({\left (}
\def\){\right )}
\def\[{\left [}
\def\]{\right ]}

\def\Nex{ N_{\text{ex}} }

\def\ii{ \mathrm{i} }

\newcommand{\beq}{\begin{equation}}
\newcommand{\eeq}{\end{equation}}

\begin{document}
\date{\today}
\flushbottom \draft
\title{Modified Bloch-Redfield Master Equation for Incoherent Excitation of Multilevel Quantum Systems}

\author{Timur V. Tscherbul}
\author{Paul Brumer}
\affiliation{Chemical Physics Theory Group, Department of Chemistry, and Center for Quantum Information and Quantum Control, University of Toronto, Toronto, Ontario, M5S 3H6, Canada}\email[]{ttscherb@chem.utoronto.ca}

\begin{abstract}
We present an efficient theoretical method for calculating the time evolution of the density matrix of a multilevel quantum system weakly interacting with incoherent light. The method combines the Bloch-Redfield theory  with a partial secular approximation for one-photon coherences,  resulting in a master equation that  explicitly exposes the reliance on transition rates and the angles between transition dipole moments in the energy basis.
The modified Bloch-Redfield master equation allows an unambiguous distinction between the regimes of quantum coherent vs. incoherent energy transfer under incoherent light illumination. The fully incoherent regime is characterized by   orthogonal transition dipole moments in the energy basis, leading to a dynamical evolution governed by a coherence-free Pauli-type master equation. The coherent regime requires non-orthogonal transition dipole moments in the energy basis, and leads to the generation of noise-induced quantum coherences and population-to-coherence couplings.
As a first application, we consider the dynamics of excited state coherences arising under incoherent  light excitation from a single ground state,  and observe population-to-coherence transfer and the formation of non-equilibrium quasisteady states   in the regime of small excited state splitting. Analytical expressions derived earlier for the V-type system [Phys. Rev. Lett. {\bf 113}, 113601 (2014)] are found to provide a nearly quantitative description of  multilevel excited-state populations and coherences in both the small- and large-molecule limits.



\end{abstract}

\maketitle
\clearpage
\newpage

\section{Introduction}


A great deal of what is known about the dynamics of photosynthetic energy transfer in higher plants and marine algae is based on the information obtained from ultrafast spectroscopic experiments on light-harvesting complexes (LHCs)   \cite{FlemingARPC,Review2,PaulLeonardoPCCP}. These experiments use femtosecond laser pulses to generate molecular excitations, which then migrate within the complex network of chromophores composing the LHCs and transfer their excitation energy to the reaction center where charge separation and subsequent steps of photosynthesis occur \cite{FlemingARPC}.  Recent experimental observations of long-lasting wavelike energy transfer in LHCs \cite{Fleming,Scholes,PaulLeonardoPCCP} has triggered a  wave of experimental and theoretical research into the possible role of quantum coherences in photosynthesis \cite{ScholesJPCL,Schulten}.


In contrast to femtosecond laser light used in the laboratory, natural photosynthesis uses incoherent light to excite the system. Incoherent light lacks the temporal and spatial coherence of a femtosecond laser pulse, and is  incident on the LHCs on a much longer timescale. As a result,  the nature of molecular states prepared by incoherent excitation can be very different from those prepared by femtosecond laser excitation \cite{JiangPaul91,MoshePaulPNAS,Kassal13,Mancal}.  In particular, it has been shown theoretically that incoherent light illumination  produces {\it incoherent mixtures of individual molecular eigenstates}  rather than their coherent superpositions.

Notwithstanding the conclusions of Refs. \citenum{JiangPaul91,MoshePaulPNAS,Mancal,Kassal13}, an entirely different kind of quantum coherence among the  excited states is known to arise  \cite{Scully92,Scully11,Scully13,Scully06,Dorfman11,Plenio,Keitel}, and persist for a long time \cite{prl} under incoherent light illumination. These coherences are due to the  
Agarwal-Fano (AF) interference between different incoherent light-induced transition pathways \cite{Scully92,Keitel,Agarwal,AgarwalBoltzmann}, and have much in common with previously explored quantum optical phenomena such as vacuum-induced coherence, lasing without inversion,  electromagnetically induced transparency, and quantum control \cite{Keitel,prl,QCbook}. This kind of noise-induced coherence has recently attracted much interest in view of its potential to  enhance the efficiency of quantum heat engines, both natural and artificially designed  \cite{Scully11,Scully13,Dorfman11}.

Despite their obvious relevance to photosynthetic light harvesting and photovoltaics,  conditions for the generation and preservation of the Agarwal-Fano (AF) noise-induced coherences in multilevel quantum systems remain unexplored. To our knowledge there have been no theoretical studies of these coherences in systems with more than a few excited levels. Even for  the model three and four-level  systems, previous theoretical work has been largely limited to the steady-state analysis with invoked drastic approximations, such as the degeneracy of the excited levels \cite{Scully06,Scully11}. A notable exception is the work of Hegerfeldt and Plenio \cite{Plenio}, which however focused on the properties of light emitted by the three-level system, rather than the system itself.  We have recently shown \cite{prl} that incoherent light can induce long-lived excited state coherences in a model V-type system provided that the excited state level splitting is  smaller than their radiative linewidths, and the dipole moments for the radiative transitions are parallel. However, real molecular systems are comprised of hundreds and thousands of excited vibronic levels, making it necessary to accurately simulate the dynamics of multilevel quantum systems excited by weak incoherent radiation.

The  Bloch-Redfield (BR) master equation formalism  \cite{CTbook} provides a general framework for such simulations. These equations describe both the incoherent pumping and spontaneous decay of quantum systems with arbitrary level structure. The only approximation involved is the Born-Markov approximation, requiring that the coupling to radiation is weak, and that the memory time of the radiation is infinitely short. As is well known \cite{BPbook}, this approximation is well justified for atomic and molecular systems pumped by weak incoherent radiation. Recent theoretical work \cite{Cao} explored the performance of BR equations against more sophisticated (and realistic) theoretical methods based on the hierarchy equations of motion \cite{IF} and found that  BR theory provides an accurate description of weak-field incoherent excitation of a chromophoric dimer in all parameter ranges \cite{Cao}. When properly applied, the BR theory can be a powerful theoretical tool for simulating excited state dynamics in molecular aggregates coupled to a surrounding protein environment, even beyond the weak-coupling regime \cite{PlenioBR}.


The BR equations have been applied to study the dynamics of incoherent excitation of few-level atomic systems in quantum optics  \cite{Scully92,Scully06,Scully13,Keitel,Scully11}. In these studies this approach is typically derived ``from scratch'' in operator form, which is advantageous for exploring the mathematical properties and  physical meaning of the  terms in the master equation \cite{SZbook,Scully11}.  Here we take an alternative approach, starting from the BR equations in the energy basis \cite{CTbook},  and apply a partial secular approximation that allows us to  simplify the equations. The resultant  modified BR equations have a clear physical meaning and are parametrized by experimentally observable quantities such as the spontaneous decay and incoherent pumping rates, and show that the alignment parameters between different transition dipole moments play a significant role. They  allow for a clear and unambiguous distinction  between the effects of quantum coherent vs. incoherent energy transfer under incoherent light illumination. Specifically, the incoherent regime is characterized  by orthogonal transition dipole moments in the energy basis. In this regime, the time evolution of the system's reduced density matrix is shown below to be  governed by the standard Pauli-type rate equations, and  a fully incoherent mixture of energy eigenstates is immediately formed, whose character ({\it e.g.} localization properties) determines the photoreaction efficiency or quantum yield \cite{jpca14}.   In the opposite regime of non-orthogonal transition dipoles, we observe nontrivial dynamical effects due to the Agarwal-Fano interference \cite{prl}, prior to long time relaxation to energy eigenstates.




The paper is organized as follows. Section II presents the derivation of the BR equations and applies the partial secular approximation to arrive at a modified system of  equations. The equations are then specialized to the case of a single-ground state and multiple excited-state systems in Sec. IIIB. Numerical results for the population and coherence dynamics in model 3, 6, and 11-level systems are presented and discussed in Sec. III. The paper concludes with a brief summary of main results and a range of open questions to be addressed (Sec.~IV).


\section{Theory}

\subsection{Bloch-Redfield equations for incoherent excitation of multilevel systems: General theory}

In this section we derive the Bloch-Redfield equations of motion for incoherent excitation of a multilevel quantum system comprising of two manifolds separated by an energy gap $\hbar\omega_0$, as shown in Fig.~1.  The combined system plus radiation field Hamiltonian is
\begin{equation}\label{H}
\hat{H} = \hat{H}_S + \hat{H}_B + \hat{V}_{SB},
 \end{equation}
where $\hat{H}_S |i\rangle = \sum_i \epsilon_i |i\rangle \langle i|$ is the system Hamiltonian with the eigenvalues $\epsilon_i$ and eigenstates $|i\rangle$.  $\hat{H}_B = \sum_{k\lambda}\omega_k \hat{a}^\dag_{\mathbf{k}\lambda} \hat{a}_{\mathbf{k}\lambda}$ is the Hamiltonian of the bath, representing a collection of non-interacting field modes with wavevectors $\mathbf{k}$ and polarizations $\lambda=1,2$, created and destroyed by the operators $\hat{a}^\dag_{\mathbf{k}\lambda}$ (in the following we set $\hbar =1)$. Incoherent thermal radiation is characterized  by the canonical density operator $\rho_B = e^{-\beta H_B}/\text{Tr}_Be^{-\beta H_B}$ where $\beta=1/k_BT_S$ is the inverse temperature of the radiation source (for the Sun, $1/\beta = 0.5$ eV at $T = 5800$~K), and $\text{Tr}_B$ denotes trace over the bath degrees of freedom.

In the dipole approximation, the system-bath coupling in Eq. (\ref{H}) takes the form \cite{BPbook,SZbook}
\begin{equation}\label{HSB}
\hat{V}_{SB} = -\hat{\bm{\mu}} \sum_{\mathbf{k}, \lambda} \left( \frac{\hbar \omega_k}{2\epsilon_0 V} \right)^{1/2}  \bm{\epsilon}_{\mathbf{k}\lambda} ( \hat{a}_{\mathbf{k}\lambda} - \hat{a}^\dag_{\mathbf{k}\lambda} )
\end{equation}
where $\hat{\bm{\mu}}$ is the transition dipole moment operator of the molecular system, and $\hat{a}^\dag_{\mathbf{k}\lambda}$ creates a photon with wavevector $\mathbf{k}$,  polarization $ \bm{\epsilon}_{\mathbf{k}\lambda}$, and frequency $\omega_k$. Expanding the transition dipole moment operator in system eigenstates $|i\rangle$ with energies $\epsilon_i$ and transforming Eq. (\ref{HSB}) to the interaction picture, we find
\begin{equation}\label{HSB2}
\hat{V}_{SB}(t) = -\sum_{i, j} \sum_{\mathbf{k}} g_{\mathbf{k}}^{(i,j)} e^{\ii (\omega_{ij} - \nu_k)t} |i\rangle \langle j| \hat{a}_\mathbf{k} + \text{H.c.}
\end{equation}
where the light-matter coupling coefficients are given by $g_{\mathbf{k}}^{(i,j)} = \bm{\mu}_{i j} \cdot \bm{\epsilon}_{\mathbf{k}\lambda}$ with $\bm{\mu}_{ij} = \langle i |\hat{\bm{\mu}}|j\rangle$ being the matrix elements of the transition dipole moment operator in the energy basis \cite{SZbook}.
The first term on the right-hand side of Eq. (\ref{HSB2}) corresponds to an absorption transition between states $|i\rangle$ and $|j\rangle$ whereby a  photon is destroyed by the operator $\hat{a}_\mathbf{k}$. Spontaneous and stimulated emission transitions are contained by the Hermitian conjugate (H.c.) term in Eq. (\ref{HSB2}).

Substituting Eq. (\ref{HSB}) into the Liouville-von Neumann equation for the total (system-plus-bath) density matrix, tracing over the bath degrees of freedom to obtain the reduced density matrix of the system, and invoking the Born-Markov approximation \cite{SZbook,BPbook} gives the integro-differential equation \cite{Blum,SZbook}
\begin{equation}\label{Leq}
\dot{\hat{\rho}}(t) = -{\ii}\text{Tr}_B[\hat{V}(t), \hat{\rho}(t_0)\otimes \hat{\rho}_B(t_0)]  - \text{Tr}_B \int_{t_0}^t [\hat{V}(t), [\hat{V}(t'),\hat{\rho}(t')\otimes\hat{\rho}_B(t_0)]] dt'.
\end{equation}
where the system-plus-bath density matrix is assumed to be a tensor product $\hat{\rho}(t')\otimes\hat{\rho}_B$ at all times, and the density matrix of the bath $\rho_B$ corresponds to a canonical distribution (see above).  Since Eq.  (\ref{HSB})   is time independent, it, in conjunction with Eq. (\ref{Leq}), implies sudden turn-on of the incoherent radiation at $t=t_0$.       To derive a master equation for the reduced density matrix of the system, we follow the standard procedure \cite{BPbook, SZbook,Blum}, and  substitute Eq. (\ref{HSB2}) into Eq. (\ref{Leq}). The second term on the right-hand side of Eq. (\ref{Leq}) gives a sum of products of interaction-picture system-bath coupling operators at two times ($V_{SB}(t) V_{SB}(t')$) and the system density matrix at time $t$. A typical term in the sum has the form
\begin{equation}\label{Vproducts}
\sum_{i j} \sum_{k l}  \sum_{\mathbf{k}} g_{\mathbf{k}}^{(i,j)} g_{\mathbf{k}}^{(k,l)} \langle \hat{a}_\mathbf{k} \hat{a}^\dag_\mathbf{k}\rangle
|i\rangle \langle j| \rho(t) |k\rangle \langle l|
\end{equation}
where $\langle \ldots \rangle = \text{Tr}_B (\ldots \rho_B)$ denotes the canonical average. Note  that the system-radiation field coupling enters all the terms in the master equation via a common prefactor
\begin{equation}\label{Vproducts2}
\sum_{i, j, k, l}   \sum_{\mathbf{k}} g_{\mathbf{k}}^{(i,j)} g_{\mathbf{k}}^{(k,l)} = \sum_{i, j,k, l} \sum_{\mathbf{k}}  ( \bm{\mu}_{i j} \cdot \bm{\epsilon}_{\mathbf{k}\lambda} )
(\bm{\mu}_{kl} \cdot \bm{\epsilon}_{\mathbf{k}\lambda}  )
\end{equation}
The exact form of this coupling depends on the nature of the incoherent radiation field. Here we are interested in the excitation of atomic and molecular systems with isotropic and unpolarized blackbody radiation, in which case the sum over $\mathbf{k}$ in Eq. (\ref{Vproducts2})  can be evaluated to yield  \cite{Agarwal, SZbook}
 \begin{equation}\label{Vproducts3}
\sum_{\mathbf{k}}  ( \bm{\mu}_{i j} \cdot \bm{\epsilon}_{\mathbf{k}\lambda} ) (\bm{\mu}_{kl} \cdot \bm{\epsilon}_{\mathbf{k}\lambda}  ) =  \bm{\mu}_{i j} \cdot  \bm{\mu}_{k l}
\end{equation}

Combining Eqs. (\ref{Leq}) --  (\ref{Vproducts3}) leads to a set of Bloch-Redfield equations in the interaction picture
\begin{equation}\label{BReq}
\dot{\rho}^I_{ij}(t) = \sum_{k,l} R_{ijkl} e^{\ii(\omega_{ij}-\omega_{kl})t} \rho^I_{kl}(t),
 \end{equation}
 where $R_{ijkl}$ are the elements of the relaxation tensor and $\omega_{ij}=\epsilon_i-\epsilon_j$ is the energy gap between eigenstates $|i\rangle$ and $|j\rangle$.
An equivalent form of the BR equations for the reduced density matrix  in the Schr{\"o}dinger picture $\rho_{ij}(t)$ may be obtained by setting $\rho_{ij}(t) =e^{-\ii\omega_{ij}t} \rho^I(t)$ into Eq. (\ref{BReq})  \cite{PFreview, PF94}
\begin{equation}\label{BReqSP}
\dot{\rho}_{ij}(t) = - \ii \omega_{ij} \rho_{ij}(t) + \sum_{k,l} R_{ijkl} \rho_{kl} (t)
 \end{equation}
The elements of the relaxation tensor in the eigenstate basis are given by \cite{CTbook}
\begin{equation}\label{R}
R_{ijkl} = -\delta_{jl} \sum_r  \Gamma^+_{irrk} + \Gamma^+_{ljik} + \Gamma^-_{ljik} - \delta_{ik}\sum_r \Gamma^-_{lrrj}
 \end{equation}
where the summations extend over all system eigenstates. The four-index correlation tensors  $\Gamma^\pm_{irrk}$  describe the coupling between the multilevel system and the isotropic and unpolarized blackbody radiation \cite{CTbook}.  In general, the tensor elements $ \Gamma^+_{ijkl}$  are complex;  the imaginary part of $ \Gamma^\pm_{ijkl}$ contributes to the coherent evolution of the system (the first term on the right-hand side of Eq. (\ref{BReqSP})).  Here, we neglect the Lamb shift contribution,  expected to be negligible for weak system-radiation couplings.  The real part of $ \Gamma^+_{ijkl} $ can be evaluated from Eqs. (\ref{Leq}) -- (\ref{Vproducts3}) following the standard procedure \cite{SZbook,CTbook} and is given by \cite{Plenio}
\begin{align}\label{GammaPlus}\notag
 \Gamma^+_{ijkl} &= \frac{1}{\epsilon_0 \hbar (2\pi c)^3} \frac{4\pi^2}{3} (\bm{\mu}_{ij} \cdot \bm{\mu}_{kl}) \, \omega_{kl}^3 \bar{n}(\omega_{kl}) \quad (\omega_{kl}>0); \\
{} &= \frac{1}{\epsilon_0 \hbar (2\pi c)^3} \frac{4\pi^2}{3} (\bm{\mu}_{ij} \cdot \bm{\mu}_{kl}) \omega_{kl}^3 \, [1+ \bar{n}(\omega_{kl}) ] \quad (\omega_{kl}<0),
 \end{align}
where $\bar{n}(\omega) = [1-e^{\omega/k_BT}]^{-1}$ is the thermal occupation number at frequency $\omega$ and temperature $T$, and $\bm{\mu}_{ij} = \langle i | \hat{\bm{\mu}} |j\rangle$ are the matrix elements of the transition dipole moment in the energy basis. The other matrix elements in Eq. (\ref{R}) can be obtained from Eq. (\ref{GammaPlus}) via the identity $\Gamma^{-}_{lkji}=(\Gamma^{+}_{ijkl})^*$. An important feature of Eq. (\ref{GammaPlus}) is the presence of the scalar product of two transition dipole matrix elements  $(\bm{\mu}_{ij} \cdot \bm{\mu}_{kl})$. This scalar product does not depend on the polarization vector $\bm{\epsilon}_\mathbf{k}$  as expected from Eq. (\ref{Vproducts3}),  reflecting the isotropic and unpolarized incident incoherent radiation.

It is instructive to define the alignment parameter  between the transition dipole moment vectors
\begin{equation}\label{p}
p_{ij,kl} = \cos\theta_{ij,kl} = \frac{\bm{\mu}_{ij} \cdot \bm{\mu}_{kl}}{\mu_{ij} \mu_{kl}}
 \end{equation}
 which can take on values from 1 (parallel transition dipoles) to -1 (antiparallel transition dipoles). As shown below, this parameter ``switches on'' the couplings between the populations and coherences, and thus plays a fundamental  role  in the dynamics of noise-induced quantum coherences induced by incoherent light illumination of multilevel molecular systems.

Using Eq. (\ref{p}) along with the standard definitions of spontaneous emission and incoherent pumping rates for the $i\to j$ transition
\begin{align}\label{rates}\notag
r_{ij} &= \gamma_{ij} \bar{n}(\omega_{ij}),\\
\gamma_{ij}&= \frac{1}{\epsilon_0 \hbar (2\pi c)^3} \frac{4\pi^2}{3} \mu_{ij}^2 \, \omega_{ij}^3,
 \end{align}
 Eqs. (\ref{GammaPlus})  can be recast in a more physically transparent form
\begin{align}\label{GammaPlus2}\notag
 \Gamma^+_{ijkl} &= \frac{1}{2} p_{ij,kl} \sqrt{r_{ij}r_{kl}} \qquad \quad\qquad\qquad (\omega_{kl}>0), \\
  \Gamma^+_{ijkl} &= \frac{1}{2} p_{ij,kl} \left[ \sqrt{r_{ij}r_{kl}} + \sqrt{\gamma_{ij}\gamma_{kl}} \right] \quad \,\,\, (\omega_{kl}<0).
 \end{align}
This expression shows that the spectral correlation tensor element $ \Gamma^+_{ijkl}$ can be expressed via sums of products of  spontaneous  decay and incoherent pumping rates. The expressions for $\omega_{kl}<0$ contain terms proportional to $\sqrt{\gamma_{ij}}\sqrt{\gamma_{kl}}$, which represents the effects of spontaneous emission. As demonstrated below, the terms with $p_{ij,kl}\ne 0$ lead to quantum interference between incoherent pumping and decay pathways  $i\to j$ and $k\to l$  \cite{Agarwal,Keitel}, and play  a crucial role in the generation of noise-induced coherences.

At this point, we restrict attention to a  system spectrum consisting of two sets of eigenstates separated by an energy gap $\omega_0$ that is large compared to the splitting between the states within each manifold (see Fig. 1).  This motivates splitting  the summation in Eq. (\ref{BReqSP})  into four terms:
\begin{multline}\label{BReqSPsplit}
\dot{\rho}^I_{ij}(t) =  \sum_{g_k, g_l} R_{ij g_k g_l}e^{\ii (\omega_{ij}-\omega_{g_kg_l})t}\rho^I_{g_k g_l}(t) + \sum_{g_k, e_l} R_{ij g_k e_l}  e^{\ii (\omega_{ij}-\omega_{g_k e_l})t} \rho_{g_k e_l}^I \\ +\sum_{e_k, g_l} R_{ij e_k g_l}  \rho_{e_k g_l}^I (t) +  \sum_{e_k, e_l} R_{ij e_k e_l}e^{\ii (\omega_{ij}-\omega_{e_k e_l})t}\rho^I_{e_k e_l}(t)
 \end{multline}
where $g_k$  and $e_k$ enumerate the states in the ground and excited-state manifolds (from now on  Latin letters are used to index the states that belong to both  manifolds). We are interested in the populations and coherences between the levels in the ground and excited-state manifolds.

To derive an equation of motion for excited state coherences $\rho_{e_i e_j}(t)$, we substitute $i\to e_i$  and $j\to e_j$ on the left-hand side of Eq. (\ref{BReqSP}) and observe that $\omega_{g_k,e_l}\sim \omega_0 \gg \omega_{e_ie_j}$. Hence, the second and third terms on the right-hand side of Eq. (\ref{BReqSP}) oscillate via the complex exponential factor $e^{\ii \omega_0 t}$, at  close to the optical frequency $\hbar\omega_0 \sim 10^{15}$ Hz. As a result, we can apply the secular approximation  to the optical transitions between the states of different manifolds.   Transforming Eq.~(\ref{BReqSP})  back to the Schr{\"o}dinger picture, and using Eq. (\ref{R}) gives
\begin{align}\label{rhoex}\notag
\dot{\rho}_{e_ie_j}(t) = -\ii\omega_{e_ie_j} \rho_{e_ie_j}(t) &+ \sum_{g_k, g_l} \left[ \Gamma^+_{g_le_je_ig_k} + \Gamma^-_{g_le_je_ig_k}  \right]
\\ &-\sum_{e_k} \rho_{e_k e_j} \sum_{g_r} \Gamma^+_{e_i g_r g_r e_k}  - \sum_{e_k} \rho_{e_i e_l} \sum_{g_r} \Gamma^-_{e_l g_r g_r e_i}.
 \end{align}
Here, we made use of the fact that $\langle e_i |\hat{\bm{\mu}}| e_j\rangle = \langle g_i |\hat{\bm{\mu}}| g_j\rangle = 0$ for any $i,j$,  so according to Eq.~(\ref{GammaPlus}) or (\ref{GammaPlus2}) the tensor elements $\Gamma^\pm_{e_irrg_k}$ vanish identically.
A similar equation of motion may be obtained for the ground-state manifold by replacing, in Eq. (\ref{rhoex}),  every index $e_i$ by $g_i$ and vice versa.


To simplify Eq. (\ref{rhoex}) we express the tensor matrix elements $\Gamma^+_{ijkl}$ in terms of incoherent pumping and spontaneous emission rates using Eq.  (\ref{GammaPlus2}). The  resulting equations of motion  take the form
\begin{align}\label{rhoex2}\notag
\dot{\rho}_{e_ie_j}(t) &= -\ii \omega_{e_ie_j} \rho_{e_ie_j}(t) + \sum_{g_k, g_l}  p_{g_l e_j, e_i g_k}   \sqrt{r_{g_ke_i}r_{g_l e_j}}   \rho_{g_k g_l}(t)
\\&- \frac{1}{2} \sum_{e_k}\rho_{e_k e_j} \sum_{g_r}   p_{e_i g_r, e_k g_r}\left( \sqrt{\gamma_{e_i g_r}\gamma_{e_k g_r}} + \sqrt{r_{e_ig_r}r_{e_k g_r}} \right) \notag
\\ &- \frac{1}{2} \sum_{e_k}\rho_{e_i e_k} \sum_{g_r}  p_{e_k g_r, e_j g_r} \left( \sqrt{\gamma_{e_k g_r}\gamma_{e_j g_r}} + \sqrt{r_{e_k g_r}r_{e_j g_r}} \right)
 \end{align}
These equations describe the  generation and destruction of quantum coherences under incoherent light excitation of a multilevel quantum system, and form the central result of this work.

Consider the physical significance of the various terms on the right-hand side. The term $\ii \omega_{e_ie_j} \rho_{e_ie_j}(t)$ describes the unitary evolution of the quantum system under the Hamiltonian $\hat{H}_S$ 
in the absence of dissipation and decoherence. The second term describes the generation of coherences due to incoherent pumping from the ground-state manifold. The coherences arise as a result of the interaction of excited states with the same field mode, a phenomenon  predicted to occur in a V-type system \cite{Dorfman11,Plenio,Keitel}. Below, we show that Eq. (\ref{rhoex2}) reduces to the equations of motion previously derived  for three-level $\Lambda$ and V-type systems\cite{Dorfman11,Plenio}.
The advantage of Eq. (\ref{rhoex2}) is its generality;  it applies to a multilevel system with an arbitrary number of ground and excited-state levels.


The last two terms in Eq. (\ref{rhoex2}) describe the decay of the   coherences via spontaneous and stimulated emission; spontaneous  decay via the terms proportional to $\sqrt{\gamma_{e_i g_r}\gamma_{e_k g_r}}$ and stimulated emission   by the terms proportional to $\sqrt{r_{e_i g_r}r_{e_k g_r}}$. Both of these terms are important for isolated atoms and molecules, where they give rise to {\it e.g.}, vacuum-induced coherence in spontaneous emission \cite{Agarwal}, which has recently been observed experimentally in quantum dots \cite{QD05}. For a molecule in the presence of an external environment, these vacuum effects are, as we have shown,\cite{prl}      of limited relevance since isolated resonances are subject to rapid environmental relaxation,  transferring  populations away from the excited eigenstates  and limiting  coherence lifetimes.  The effects of environmentally-induced relaxation and decoherence  can be included in Eq. (\ref{rhoex2}) by adding system-specific terms that describe the coupling to a phonon bath \cite{prl}.

The alignment parameters in Eq. (\ref{rhoex2}) form a square transition dipole alignment matrix with elements $p_{ij}$. For  $p_{e_i g_r, e_k g_r} = \delta_{e_i, e_k}$  the time evolution of the populations is decoupled  from that of the coherences, leading to Pauli-type rate equations for the populations and the exponential decay of the coherences \cite{CTbook,SZbook,BPbook}. A diagonal  $p_{ij}$  results whenever the transition dipole moments for the transitions $g_r \to e_i$ and $g_r \to e_k$ are orthogonal.
In the case of incoherent excitation from the ground state $\rho_{gg}=1$ (where all other elements of the density matrix are zero),  it follows from Eqs. (\ref{rhoex2}) that there are no coherences  generated by incoherent excitation if the transition dipole moments are all orthogonal to one another, that is $p_{ij}=\delta_{ij}$.



\subsection{Application to V- and $\Lambda$-type systems}

Here, we specialize the BR equations of motion (\ref{rhoex2}) to the case of incoherent excitation to a manifold of states from a single ground level. This model provides a reasonable starting point for the description of incoherent excitation of  internally cold atoms, molecules  or quantum dots \cite{Scully06}, in which thermal excitations from the absolute ground state can be neglected.

For the ground-state manifold consisting of a single state $|g\rangle$, Eqs. (\ref{rhoex2})  to yield
\begin{align}\label{rhoex3}\notag
\dot{\rho}_{ij}(t) = -\ii \omega_{ij} \rho_{ij}(t) +  p_{ij} \rho_{gg}(t) \sqrt{r_{i}r_{j}}
 &-  \frac{1}{2}\sum_{k}\rho_{kj}(t) p_{ik}  \left( \sqrt{\gamma_{i}\gamma_{k}} + \sqrt{r_{i}r_{k}} \right)  \\
&-  \frac{1}{2}\sum_{k}\rho_{ik}(t) p_{jk}   \left( \sqrt{\gamma_{j}\gamma_{k}} + \sqrt{r_{j}r_{k}} \right)
 \end{align}
where denote $i=e_i$, $j=e_j$, $\gamma_{gi} = \gamma_i$, and   $p_{ge_i,ge_j}=p_{ij}$. Note that trace conservation implies $\rho_{gg}=1-\sum_{k}\rho_{kk}$, so Eq. (\ref{rhoex3}) can be formulated entirely in terms of the excited eigenstates. In addition, Eq. (\ref{rhoex3})  establishes that the time propagation of any given density matrix element $\rho_{ij} $ is determined by the matrix elements $\rho_{kj}$ in the same row (or column) of the excited-state density matrix. The other columns are decoupled, a feature which may be used to design more  efficient algorithms for the numerical solution of the equations of motion. Even without  taking advantage of this property, numerical solution of Eqs. (\ref{rhoex3}) is more straightforward and computationally efficient than direct integration of the BR equations (\ref{BReqSP}). This is because only two matrix-matrix products are required on each time step for Eq. (\ref{rhoex3}), as compared to more than ten matrix operations in the original $O(N^3)$ algorithm of, {\it e.g.},  Pollard and Friesner  \cite{PF94,PFreview}.


 Note that in the case where  excitation occurs only to a pair of excited eigenstates $|1\rangle$ and $|2\rangle$,  the equations of motion for the  coherence reduce to
\begin{equation}\label{rhoex4}
\dot{\rho}_{12}(t) = -\frac{1}{2}\left( r_1+r_2 + \gamma_1 + \gamma_2 \right) +  p\sqrt{r_1 r_2} \rho_{gg} -  \frac{1}{2}p ( \sqrt{r_1 r_2} +\sqrt{\gamma_1\gamma_2}) (\rho_{11}+\rho_{22}) - \ii \omega_{12}\rho_{12}
 \end{equation}
 where $p=p_{12}$. This is the equation of motion for the V-type system pumped by incoherent radiation \cite{Scully06,Scully92,prl}, demonstrating that the modified BR equations for  excitation from a single ground state  reduce to the correct expressions in the limit of one ground and two excited states.

For the sake of completeness, we also consider the decay of a single excited state $|e\rangle$ to a manifold of ground-state levels $|g_i\rangle = |i\rangle$. This decay is known to generate coherence between the $|g_i\rangle$ provided the $e\to g_i$  transition dipole moments are not orthogonal \cite{AgarwalLambda}, and has been observed experimentally in GaAs quantum dots at very low temperatures \cite{QD05}. The equation of motion for $\rho$ follows from Eq. (\ref{rhoex2})  in the single excited-state limit:
\begin{align}\label{rhoexLambdaMany}\notag
\dot{\rho}_{ij}(t) = -\ii \omega_{ij} \rho_{ij}(t) + \rho_{ee}(t) (\sqrt{r_{i}r_{j}}    + \sqrt{\gamma_{i}\gamma_{j}} )
 &-  \frac{1}{2} \sum_{k}\rho_{kj} p_{ik}  \sqrt{r_{i}r_{k}}  \\
&- \frac{1}{2}\sum_{k}\rho_{ik} p_{jk}  \sqrt{r_{j}r_{k}}~,
 \end{align}
where the sums over $k$ do not include the excited state. If the excited state decays into two ground states $|1\rangle$ and $|2\rangle$, for example, Eqs. (\ref{rhoexLambdaMany}) reduce to
\begin{equation}\label{rhoexLambda}
\dot{\rho}_{12}(t) = -\frac{1}{2}\left( r_1+r_2 \right) +  p (\sqrt{r_1 r_2} + \sqrt{\gamma_1\gamma_2}) \rho_{ee} -  \frac{1}{2}p  \sqrt{r_1 r_2}  (\rho_{11}+\rho_{22}) - \ii \omega_{12}\rho_{12}
 \end{equation}
which is the equation of motion for the ground-state coherence in the $\Lambda$-type system \cite{Ou,Plenio}.

\section{Results and Discussion}

\begin{figure}[t!]
	\centering
    \includegraphics[width=0.6\textwidth, trim = 0 0 0 0]{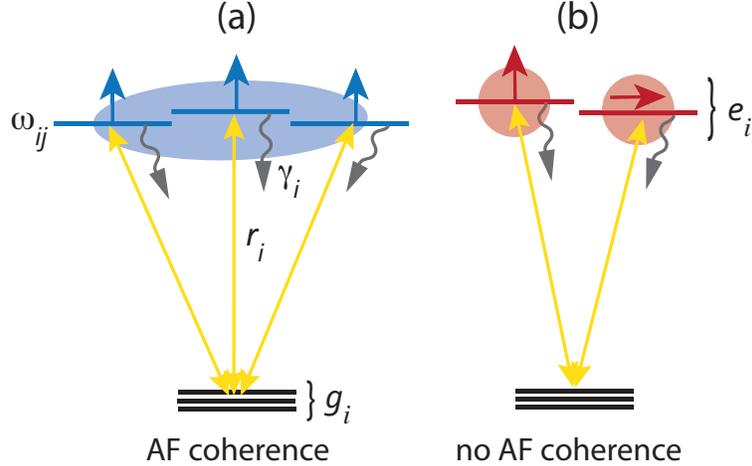}
	\renewcommand{\figurename}{Fig.}
	\caption{Incoherent excitation of a multilevel quantum system with parallel (a) and orthogonal (b) transition dipole moments. Double-sided  arrows represent incoherent pumping and stimulated emission transitions, one-sided arrows indicate the orientation of transition dipole moments. Shaded areas represent the coherences generated by incoherent pumping among the excited eigenstates.}
\end{figure}

As an application of the modified BR theory  consider the dynamics of  a multilevel system with a single ground state illuminated by incoherent light (see Fig. 1). The single-ground-state approximation is  reasonable  for the excitation of 
cold molecules on time scales short compared to the spontaneous emission, when the excited-state population is small, and the effects of spontaneous decay back to ground state levels can be neglected (typically, less than 1 nanosecond).

Consider incoherent excitation of a model molecular system initially in the  ground state ($\rho_{gg}=1$) by isotropic and unpolarized incoherent radiation (e.g., sunlight) suddenly turned on at time zero. 
For simplicity,  we assume that (i)  $r_i=r$ and $\gamma_i=\gamma$ with $\gamma/2\pi=1$ GHz, corresponding to a typical radiative lifetime of 1 ns, (ii) the states are equidistant, with $\omega_{i,i-1}=\Delta$, and 
(iii) all transitions have dipole moments either all aligned along the same axis ($p_i=1$) or completely orthogonal ($p_i=0$). These assumptions will be relaxed in our subsequent calculations \cite{TBP}, in which a realistic set of parameters $r_i$ and $\gamma_i$ will be utilized.

\subsection{Small-molecule limit: $\Delta/\gamma\gg 1$}

Consider  first  the regime of large splitting $\Delta$ between
the  excited-state  energy levels, where $\Delta/\gamma \gg 1$,
corresponding  to  the  small-molecule  limit  \cite{Uzer}. The
resultant  off-diagonal elements of $\rho(t)$ are shown in Fig.
2  for $\Nex=10$ and $\Delta/\gamma=24$. The coherences exhibit
damped  oscillations with frequency set by the energy splitting
$\Delta$. Note that no coherent driving fields are present, and that 
these  oscillations  arise  due  to  the  sudden turn-on of the
interaction with the incoherent radiation field, hence the name
noise-induced coherences \cite{Scully06,Scully11}.

\begin{figure}[t!]
	\centering
    \includegraphics[width=0.6\textwidth, trim = 0 0 0 0]{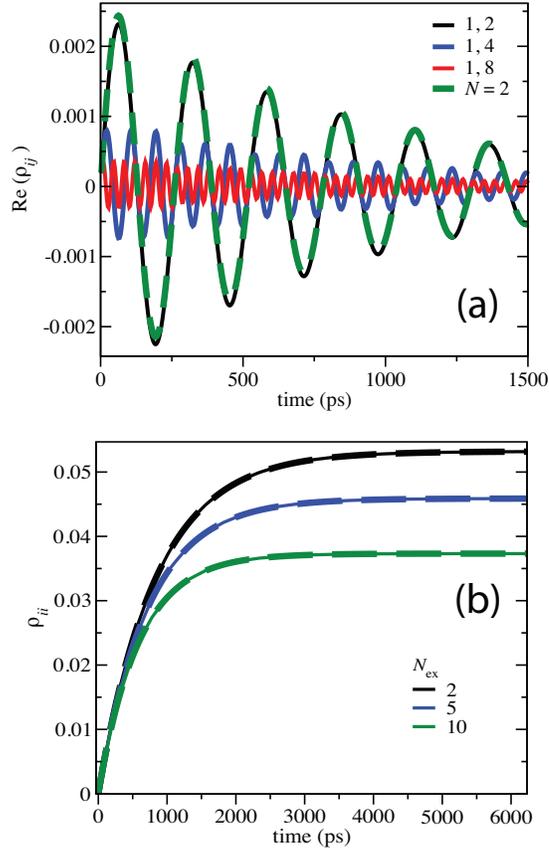}
	\renewcommand{\figurename}{Fig.}
	\caption{(a) The real part of the coherence in the energy basis: $\rho_{1,2}$ (black line), $\rho_{1,4}$, and $\rho_{1,8}$ (middle line) versus time for $\Delta/\gamma=24$, and for $p_{ij}=1$. (b) Time evolution of the excited-state populations $\rho_{11}(t)$ for different number of excited levels $\Nex=2$, 5, and 10. Results for both fully aligned ($p_{ij}=1$, full lines) and orthogonal ($p_{ij}=\delta_{ij}$, dashed lines) transition dipole moments are shown. }
\end{figure}

We  previously  showed,  for  a  V-type  system with 
$p=1$ that is weakly pumped by incoherent
light, that  the   coherence  between  the  two  upper  eigenstates
$|i\rangle$ and $|j\rangle$ evolves as \cite{prl}
\begin{equation}\label{analyticCoh}
\text{Re}(\rho_{ij}) = \left( \frac{r}{\omega_{ij}} \right) e^{-\gamma t} \sin(\omega_{ij} t),
\end{equation}
where $\omega_{12}$ was equal to $\Delta$.  
For the case of adjacent levels, the real part of the coherence is plotted as a dashed line in Fig. 2(a) , and clearly follows Eq. (\ref{analyticCoh}).    
Here we observe 
that Eq. (\ref{analyticCoh}) accurately describes the dynamics between  {\it any} two 
excited eigenstates of a multilevel V-system in the regime $\Delta/\gamma  \gg1$. Specifically,
Eq. (\ref{analyticCoh}) predicts that the maximum  of $\text{Re}\rho_{ij}$ decreases with increasing $\omega_{ij}$ while its oscillation frequency increases. This behavior is clearly visible in Fig. 2(a): The energy splitting between  eigenstates $|i\rangle$ and $|j\rangle$ is $\omega_{ij} = \Delta(j-1)$ for $j>i$, 
so, for example, the coherences $\rho_{1,j}(t)$ decrease in magnitude as $1/(j-1)$ and oscillate with frequencies $(j-1)\Delta$.
The remarkable ability of Eq. (\ref{analyticCoh}) to predict the noise-induced coherent dynamics of a multilevel system is, as we have suggested,\cite{prl} a manifestation of the fact that the isolated resonances $|i\rangle$ are not directly coupled by the incoherent light. Rather, they  experience pairwise interactions via dipole coupling to the common ground state $|g\rangle$.


Figure 2(b) displays the time dynamics of excited-state populations calculated for both parallel ($p_{ij}=1$) and orthogonal transition dipole moments ($p_{ij}=0$), which are seen to be virtually the same.  The populations smoothly evolve from zero to a steady-state value given by the 
Boltzmann distribution characterized by $\rho_{ii}/\rho_{jj}=e^{-\beta\omega_{ij}}$ for $r/\gamma = \bar{n}(\omega_0)$,
where $\omega_0$ is the splitting between the ground and excited-state manifolds of states (we take $\omega_0=1.41$~eV throughout). 
Note that the steady-state $\rho_{ii}$ values decrease with increasing  number of excited states $\Nex$; a consequence of trace conservation, since the entire population is spread over a larger number of excited-state levels.  
The analytic solution for the population of two-state system, assuming equal pumping rates,
was obtained in Ref. \citenum{prl}. Here we can extend this solution to apply, at long times, to an $N_\text{ex}$ level
system:
\begin{equation}\label{analytic_pop}
\rho_{ii} = \frac{e^{\beta\omega_0}+2}{e^{\beta\omega_0}+\Nex} \left( \frac{r}{\gamma} \right) [1 - e^{-\gamma t}]  
\end{equation}
The first factor on the right-hand side of Eq. (\ref{analytic_pop}) ensures that 
the populations tend to a canonical steady-state value at long times, {\it i.e.}, $\rho_{ii}/\rho_{gg}=e^{-\beta \omega_{0}}$ (the index $i$ is here restricted to run over the excited states). 
Significantly, in this $\Delta/\gamma \gg 1 $ regime, the behavior of the populations is independent of  transition dipole moments  alignment, in stark contrast to the coherences given by Eq. (\ref{analyticCoh}), which vanish identically for $p_{ij}=0$, as discussed at the end of Sect. IIA. 
In the V-system limit ($\Nex=2$) Eq. (\ref{analytic_pop}) reduces exactly 
to Eq. (3) of Ref.~\citenum{prl}, as expected.

The short-time behavior of both the populations and coherences shown in Fig. 2 can be reproduced by our first-order perturbation theory in light-matter interaction \cite{Zaheen,PRA}. While these perturbative equations provide additional insight into excitation dynamics at short timescales, they are fundamentally limited in two ways: first, they do not describe the approach to steady-state equilibrium at longer times. Second, the semiclassical theory of light-matter interaction does not capture the effects  of  spontaneous emission, and hence cannot be used whenever these effects are important, {\it e.g.} for strongly radiatively broadened levels and/or timescales comparable to the radiative decay time of excited-state levels $\tau=1/\gamma$.  A proper description of excitation dynamics in these regimes requires  the fully quantum BR theory  developed above.


\subsection{Large-molecule limit: $\Delta/\gamma\ll 1$}

Polyatomic  molecules  are characterized by small $\Delta/\gamma$,
with  an  average energy spacing much smaller
than  the  radiative  linewidth. For example, in a medium-sized
molecule such as pyrazine, typical rovibronic splittings are on
the   order   of  0.04  cm$^{-1}$  and  decrease  rapidly  with
increasing   molecule  size,  reaching  values  $\sim$10$^{-5}$
cm$^{-1}$  for  antracene.  This  is  to  be  compared with the
radiative decay width of $1$ GHz = 0.03 cm$^{-1}$ corresponding
to  a  radiative  decay  lifetime  of  1  ns.  It  is therefore
important   to   study   incoherent  light  excitation  in  the
``large-molecule'' \cite{Uzer} limit $\Delta/\gamma \ll 1$.

\begin{figure}[t!]
	\centering
    \includegraphics[width=0.6\textwidth, trim = 0 0 0 0]{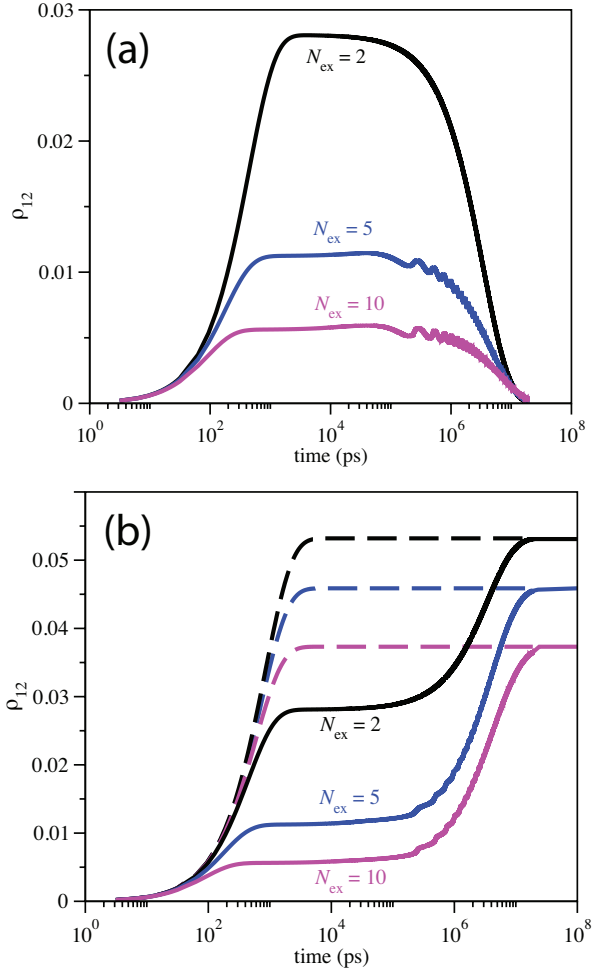}
	\renewcommand{\figurename}{Fig.}
	\caption{(a) The real part of the coherence $\rho_{1,2}$ (black line)  versus time for $\Delta/\gamma=0.024$ for different numbers of excited states $N_\text{ex}=2, 5, 10$ for $p_{ij} = 1$. (b) Time evolution of the excited-state population $\rho_{11}(t)$ for  $\Nex=2$, 5, and 10. Results $p_{ij}=1$, full lines, and $p_{ij}=\delta_{ij}$, dashed lines,  are shown. }
\end{figure}

The time dependence of the coherence between two closely spaced eigenstates is shown in Fig. 3(a) as a function of time for $\Delta/\gamma = 0.024$. Unlike the case  in the previous section, the two states now appear as radiatively broadened resonances, experiencing  decay to the continuum. The effects described below  are therefore similar to those observed in previous work on  Agarwal-Fano resonances \cite{Agarwal,Keitel}. However,  this previous work  focused on the resonances in few-level atomic systems, and was more concerned with the decay of resonances rather than with their population or coherences, the aspect that we focus on here.

\begin{figure}[t!]
	\centering
    \includegraphics[width=0.6\textwidth, trim = 0 0 0 0]{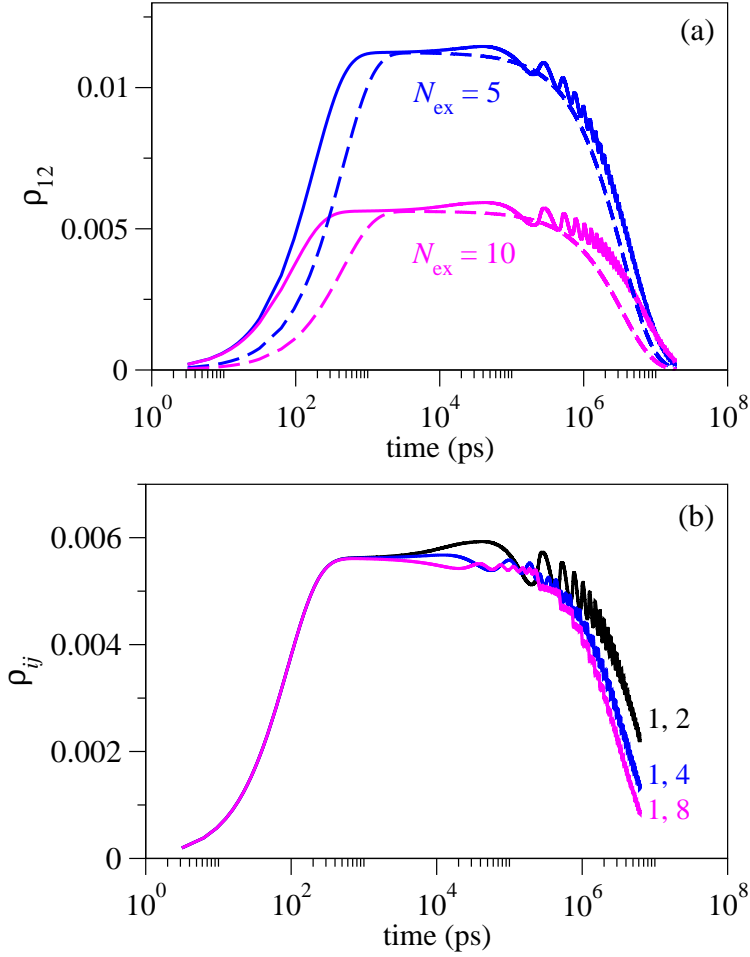}
	\renewcommand{\figurename}{Fig.}
	\caption{(a) Exact multilevel excited-state coherences $\rho_{12}(t)$ for $\Nex=5$, and 10 (full lines) modeled using Eq. (\ref{coh_smallDelta}) (dashed lines).   (b) The real part of the coherences $\rho_{1,2}$, $\rho_{1,4}$, and $\rho_{1,8}$ versus time for $\Delta/\gamma=0.024$, $N_\text{ex}=10$, and $p_{ij} = 1$.  }
\end{figure}

The most striking feature  in Fig. 3(a) is the long lifetime 
(note, here and below, the logarithmic scale of the abscissa)
of the coherences  between closely spaced eigenstates with parallel transition dipole moments.
The $\Nex=2$ case, for which the BR equations can be solved 
analytically in the weak-pump limit\cite{prl} (with $\omega_{ij} = \Delta$)
motivates consideration of the following form for coherences of a pair of levels in
the $N_{ex}$ case:
 \begin{equation}\label{coh_smallDelta}
\text{Re}\rho_{ij}(t) = \frac{2}{N_\text{ex}}\left( \frac{r}{2\gamma}\right) \left[ e^{-(\gamma/2) (\Delta/\gamma)^2 t} - e^{-2\gamma t}\right]~,  \qquad (\Delta/\gamma \ll 1)
\end{equation}
providing general insight into the decoherence dynamics.
(Note that, from Ref. \citenum{prl} one would anticipate that $\Delta$ on the right hand side of
Eq. (\ref{coh_smallDelta}) should be replaced by $\omega_{ij}$ for arbitrary levels $i$ and $j$. However,
results below in Fig. 4(b) show that Eq. (\ref{coh_smallDelta}) is correct for any
two levels that lie within the radiative line width $\gamma$ of one another).

Equation (\ref{coh_smallDelta}) indicates that
the real part of the coherence has a bi-exponential form
controlled by two timescales, the first of which is spontaneous
decay $\tau_s=1/\gamma$, which governs the time evolution of
the system at short times $t\ll \tau_s$. At $t\sim \tau_s$, the
first  exponent  on  the  right-hand  side  of  Eq.
(\ref{coh_smallDelta}) decays to zero, and the coherences
approach a {\it quasisteady state} value given by $r/(N_\text{ex}\gamma)$.
This   plateau   lasts   for   $\tau_\Delta=(2/\gamma)
(\Delta/\gamma)^{-2}$, during which the first exponent in Eq.~(\ref{coh_smallDelta})
 slowly decays to zero. Because in the
$\Delta/\gamma\gg 1$ regime $\tau_\Delta \gg \tau_s$, the
overall decoherence timescale is given by $\tau_\Delta$ and can
be arbitrarily long. It is, however, finite for $\Delta>0$, so
that   the   coherences   eventually   decay   to
zero\cite{AgarwalBoltzmann}. The prefactor $2/\Nex$ is Eq.
(\ref{coh_smallDelta})  accounts for a
decreasing amount of excited-state coherence with increasing
$\Nex$ as coherences spread over a large number of excited
states, a tendency clearly apparent in Fig.~3(a) and below.

Note that Eq. (\ref{coh_smallDelta}) is essentially an analytic
solution to the V-system problem, renormalized to
account for trace conservation in the presence of multiple
excited  states.  To  test  the accuracy  of  Eq.
(\ref{coh_smallDelta}), we plot in Fig. 4(a) the real part of
the coherence between the adjacent eigenstates in a multilevel
system as a function of time, together with the prediction of
Eq. (\ref{coh_smallDelta}) for $\Nex=5$ and 10. We observe that
the  two-level expression accurately captures the overall
effect  of  the multilevel AF coherence. However, Eq.~(\ref{coh_smallDelta})
predicts a later onset (and early
decline)  of  the quasisteady behavior compared to exact
calculations, thereby underestimating the coherence lifetime in
real multilevel systems. In addition, the analytical expression
does not reproduce the late-time oscillations present in the
exact results shown in Fig. 4(a). Nevertheless, the simple
model (\ref{coh_smallDelta}) provides a good estimate of the
magnitude of the multilevel coherences and puts a lower bound
on their lifetime in the $\Delta/\gamma\ll 1$ regime.

Figure   4(b)   shows   the   time  evolution  of  the  various
off-diagonal  elements  of  the  density  matrix  in  the small
$\Delta/\gamma$  regime  for  $\Nex=10$.  In  contrast with the
large-$\Delta$  limit for two levels, the duration of the coherences does not
decrease  with  increasing  $\omega_{ij}$,  but
rather is in accord with Eq. (\ref{coh_smallDelta}). We attribute this
remarkable and unexpected   longevity   of   multistage   coherences   to  the
constructive  AF  interference  between  three  or more excited
states,  which  cannot be properly captured by a two-level
model.  


As discussed below, the long-lived quantum coherences shown in
Fig.  3(a)  are  closely related to deviations from
thermodynamic equilibrium displayed by the populations. We
will thus refer to both of these behaviors as due to {\it
non-equilibrium quasisteady states} (NEQSs). As shown in Fig.
3(a), the coherences associated with the NEQSs form for any
value of $\Nex$ and persist for the same amount of time
$\tau_\Delta$.

Consider now the time evolution of excited-state populations. Figure~3(b) compares, as an example, the diagonal matrix elements $\rho_{11}(t)$ for $\Delta/\gamma = 0.024$ evaluated for parallel ($p_{ij}=1$, full lines) and orthogonal ($p_{ij}=0$, dashed lines ) transition dipole moments.  Figure~5 shows similar data, but for the sum of excited-state populations vs. time for different numbers of excited states $N_\text{ex}$, 
{\it i.e.} the total population excited from the ground state. As in
the $N_\text{ex}=2$ case\cite{prl}, the excited-state populations 
are  determined  by  two  independent  timescales, one of which
[$\tau_\Delta=(2/\gamma) (\Delta/\gamma)^{-2}$]  can be much longer
than  the  other [$\tau_s=1/\gamma$]. 
Unlike the coherences, however, the populations in the {\it very long-time limit}, $t\gg \tau_\Delta$ tend to their true equilibrium values set by the Boltzmann distribution, $\rho_{ii}/\rho_{gg} = e^{-\beta\omega_0}$. This long-time limit is the same as that reached in the process of incoherent excitation of the many-level system with orthogonal transition dipole moments ($p_{ij}=0$). As illustrated in Figs. 3(b) and 5, the coherence-free evolution associated with $p_{ij} = 0$, shown by the dashed lines, results in the same steady-state values of populations. An important difference is however, the time that it takes for the system to reach equilibrium; this time is much longer for the system that exhibits the Agarwal-Fano interference.

 In both Fig. 3(b) and 5, in the absence of Agarwal-Fano interference, the population is seen to  grow steadily at short times before reaching a steady state. The steady state value of the sum scales linearly with $N_\text{ex}$ as expected. The AF coherence has a profound effect on the time evolution of the population sum: as shown in Fig. 4, a quasisteady state value is reached between 100 and $10^5$ ps that does not depend on the number of excited states. This is due to the population $\to$ coherence transfer occurring on this intermediate timescale, during which the excited-state coherences increase with time (see Figs. 2 and 3).
At $t > \tau_\Delta$, the curves for different $N_\text{ex}$ start to diverge, tending to their Boltzmann steady state limits. This signals the beginning of coherence $\to$ population backtransfer, which enhances the value of the population sum for larger $N_\text{ex}$.

\begin{figure}[t!]
	\centering
  \includegraphics[width=0.5\textwidth, trim = 0 0 0 0]{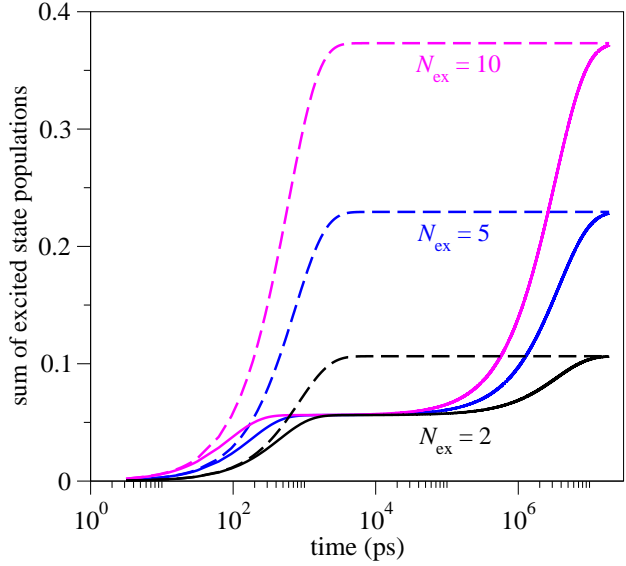}
	\renewcommand{\figurename}{Fig.}
	\caption{Sum of excited-state populations as a function of time for $\Delta/\gamma=0.024$. Results for both fully aligned ($p_{ij}=1$, full lines) and orthogonal ($p_{ij}=\delta_{ij}$, dashed lines) transition dipole moments are shown. }
\end{figure}


\section{Discussion and Summary}

This paper provides a theoretical framework for describing incoherent excitation of multilevel quantum systems by weak solar radiation, the central result being  a modified system of BR equations describing the time evolution of the system reduced density matrix. The spectrum of the solar radiation is close to that of a black body at 5800 K, with maximum intensity at transition energies  $\omega_0\sim$ 1.2-2 eV. 
Since the states within the ground and excited-state manifolds are not directly coupled by the incoherent radiation, we can apply the secular approximation to the inter-manifold transitions while retaining the essential coherences between the states within the same manifold. Invoking this approximation gives a simplified system of BR equations for the dynamics of populations and coherences under incoherent illumination. These equations are parametrized by the incoherent pumping and decay rates, as well as by the angles between the $i\to j$ and $k\to l$ transition dipole moments $p_{ij,kl}$. They can be solved by a simple and efficient $O(N^3)$ algorithm based on the modified BR equations without the need to perform numerous matrix products (as required in standard Redfield approaches \cite{PF94}), and may thus enable accurate numerical simulations of incoherent excitation of quantum multilevel systems involving hundreds to thousands of excited states.

 In the limit of orthogonal transition dipole moments, $p_{ij,kl}=\delta_{ij,kl}$, the simplified BR equations reduce to rate equations for incoherent population transfer among system eigenstates. In this limit, the time evolution of the excited-state populations is decoupled from that of the coherences and coherences are not generated by incoherent pumping from the ground state. We thus arrive at the conclusion reached in Refs. \citenum{JiangPaul91,MoshePaulPNAS} and Ref. \citenum{Kassal13}. The present work shows that this conclusion, i.e., the absence of coherences, 
 {\it is valid at all times} if the  transition dipole moments are orthogonal to one another.
In the presence of  non-orthogonal transition dipole moments, however, Agarwal-Fano coherences are initially generated, and can last for long times. Eventually, at longer times, these coherences cease to play a role and a Boltzmann steady state results, in agreement with a previous theoretical result \cite{MoshePaulPNAS}.
The timescale for the existence and role of these coherences in real biological systems (which are also subject to relaxation and decoherence due to the interaction with a protein/solvent/phonon environment) remains unexplored.

The key condition for the generation of Agarwal-Fano coherences in that of non-orthogonality of transition dipole moments: $p_{ij,kl}\ne \delta_{ij,kl}$, which give rise to couplings between the populations and coherences.  In Sec. IV we studied the noise-induced coherence dynamics in both small and large-molecule limits. In the small-molecule limit characterized by $\Delta/\gamma \gg 1$, the coherences show damped oscillations and decay to zero at times long compared to the radiative lifetime $\tau_s=1/\gamma$.

 In the large-molecule regime of small level splittings ($\Delta/\gamma \ll 1$)  extremely long-lived coherences are observed along with the formation of non-equlibrium quasi steady states. These states survive on a timescale that scales as $\Delta^{-2}$ provided that the eigenstates are close enough in energy. An analytical model recently developed for the V-type system \cite{prl} is found to apply to the multilevel case as well. With certain modifications, this model provides a nearly quantitative description of noise-induced coherences and population dynamics in the relevant regimes.
Finally, calculations show that Agarwal-Fano interferences slow down the approach to equilibrium in multilevel systems with non-orthogonal transition dipole moments. This slow approach is correlated with the presence of near equilibrium steady states characterized by finite values of the coherences [see Fig. 3(a)].

We envision several extensions of this work.
First and foremost, it is important to quantify the role of noise-induced coherences in real molecules, where several considerations are expected to be relevant: (1) intramolecular vibrational relaxation due to the presence of dark states other than the ground state, leading to secondary decay channels, (2) dipole and symmetry selection rules leading to many possible orientations of transition dipole moments, (3) the presence of incoherent transitions from multiple ground states, (4) anharmonic excited-state level spacings
and (5) possible Lamb shift effects at very small $\Delta$\cite{Altenmuller}. A first-principles simulation of dynamics of this type could be performed using the approach developed here (Eq.~\ref{rhoex2}), where the equations would be parametrized by \textit{ab initio} rovibrational energy levels and transition dipole moments. Such a dynamics calculation currently appears feasible for diatomic and modest-size polyatomic molecules, such as HCN or acetylene, which could serve as prototypes for noise-induced coherences in larger systems.


Second, for large molecules where direct eigenstate computations are not feasible, it would be necessary to develop adequate theoretical models to take into account the relaxation and decoherence of excited states induced by a phonon bath. As alluded to in Sec. IVA this task is complicated by the presence of many internal degrees of freedom and proteomic environments (such as that in {\it cis-trans} photoisomerization in retinal \cite{jpca14}) which are likely to result to rapid relaxation and decoherence of excited state coherences. However, model calculations show that these coherences can still play a non-negligible role in the energy transfer among certain eigenstates at short times, being comparable in magnitude to the populations \cite{Mancal,prl,TBP}. Theoretical work aiming to explore these effects is currently in progress.

As mentioned above, an important question that remains to be addressed is whether the multilevel Agarwal-Fano coherences considered in this work play a significant role in photosynthetic energy transfer. The mechanisms such as noise-assisted energy transfer and/or supertransfer, were proposed for the case of coherent excitation initially localized on a single chromophore \cite{Lloyd,PlenioNJP}. Under natural light illumination, the entire assembly of chromophores is excited and these mechanisms may no longer apply \cite{Kassal13}, so it is important to study the energy transfer dynamics in the light-harvesting complexes such as FMO and PC645 under incoherent illumination. Such studies are currently in progress \cite{FMO,Leonardo}.

\section*{Acknowledgements}

This work was supported by the Natural Science and Engineering Council of Canada and the U.S. Air Force Office of Scientific
Research under contract number FA9550-13-1-0005. We are grateful to Professors Gershon Kurizki and Leonardo Pach{\'o}n for stimulating discussions.


\begin{thebibliography}{99}



\bibitem{FlemingARPC}
Y. C. Cheng and G. R. Fleming, Annu. Rev. Phys. Chem. {\bf 60}, 241 (2009).

\bibitem{Review2}
V. I. Novoderezhkin and R. van Grondelle, Phys. Chem. Chem. Phys. {\bf 12}, 7352 (2010).

\bibitem{PaulLeonardoPCCP}
L. A. Pach{\'o}n and P. Brumer, Phys. Chem. Chem. Phys. {\bf 14}, 10094 (2012).


\bibitem{Fleming}
G. S. Engel, T. R. Calhoun, E. L. Read, T. K. Ahn, T. Mancal, Y. C. Cheng, R. E. Blankenship, and G. R. Fleming,
Nature (London) {\bf 446}, 782 (2007).


\bibitem{Scholes}
E. Collini, C. Y. Wong, K. E. Wilk, P. M. Curmi, P. Brumer, and G. D. Scholes, Nature (London) {\bf 463}, 644 (2010).


\bibitem{ScholesJPCL}
G. D. Scholes, J. Phys. Chem. Lett. {\bf 1}, 2 (2010).

\bibitem{Schulten}
J. Strumpfer, M. Sener, and K. Schulten, J. Phys. Chem. Lett. {\bf 3}, 536 (2012).

\bibitem{JiangPaul91}
X.-P. Jiang and P. Brumer, J. Chem. Phys. {\bf 94}, 5833 (1991); Chem. Phys. Lett. {\bf 180}, 222 (1991).

\bibitem{MoshePaulPNAS}
P. Brumer and M. Shapiro, Proc. Natl. Acad. Sci. USA {\bf 109}, 19575 (2012).

\bibitem{Mancal}
T. Man\u{c}al and L. Valkunas, New J. Phys {\bf 12}, 065044 (2010).

\bibitem{Kassal13}
I. Kassal, J. Yuen-Zhou, and S. Rahimi-Keshari, J. Phys. Chem. Lett. {\bf 4}, 362 (2013).

\bibitem{Scully92}
M. Fleischhauer, C. H. Keitel, M. O. Scully, and C. Su, Opt. Commun. {\bf 87}, 109 (1992).

\bibitem{Scully11}
M. O. Scully, K. R. Chapin, K. E. Dorfman, M. B. Kim, and A. Svidzinsky, Proc. Natl. Acad. Sci. USA {\bf 108}, 15097 (2011).

\bibitem{Scully13}
K. E. Dorfman, D. V. Voronine, S. Mukamel, and M. O. Scully, Proc. Natl. Acad. Sci. USA {\bf 110}, 2746 (2013).

\bibitem{Scully06}
V. V. Kozlov, Y. Rostovtsev, and M. O. Scully, \pra{74}{063829}{2006}.

\bibitem{Dorfman11}
A. A. Svidzinsky, K. E. Dorfman, and M. O. Scully, \pra{84}{053818}{2011}.

\bibitem{Plenio}
G. C. Hegerfeldt and M. B. Plenio, \pra{47}{2186}{1993}.

\bibitem{Keitel}
M. Kiffner, M. Macovei, J. Evers, and C. H. Keitel, Progress in Optics {\bf 55}, 85 (2010), section 3.2.5.

\bibitem{prl}
T. V. Tscherbul and P. Brumer, \prl{113}{113601}{2014}, and extensive supplementary material therein.

\bibitem{Agarwal}
A. K. Patnaik and G. S. Agarwal, \pra{59}{3015}{1999}.

\bibitem{AgarwalBoltzmann}
G. S. Agarwal and S. Menon, \pra{63}{023818}{2001}.


\bibitem{QCbook}
M. Shapiro and P. Brumer, {\it Quantum Control of Molecular Processes} (Wiley-VCH, Weinheim, 2012).

\bibitem{CTbook}
C. Cohen-Tannoudji, J. Dupont-Roc, and G. Grynberg, {\it Atom - Photon Interactions: Basic Process and Appilcations} (Wiley-VCH, Weinheim, 2004).

\bibitem{BPbook}
H.-P. Breuer and F. Petruccione, {\it The Theory of Open Quantum Systems} (Clarendon Press, Oxford, 2006), Chap. 3.4.

\bibitem{Cao}
J. Ol\u{s}ina, A. G. Dijkstra, C. Wang, and J. Cao, arXiv:1408.5385v1 (2014).

\bibitem{IF}
A. Ishizaki and G. R. Fleming, \jcp{130}{234111}{2009}.

\bibitem{PlenioBR}
J. Jeske, D. Ing, M. B. Plenio, S. F. Huelga, and J. H. Cole, arXiv:1408.2726v1 (2014).


\bibitem{SZbook}
M. O. Scully and M. S. Zubairy, {\it Quantum Optics} (Cambridge University Press, Cambridge, UK, 1997).

\bibitem{jpca14}
T. V. Tscherbul and P. Brumer, J. Phys. Chem. A {\bf 118}, 3100 (2014).



\bibitem{Blum}
K. Blum, {\it Density Matrix Theory and Applications} (Springer-Verlag Berlin, 2012), Chap. 8.


\bibitem{QD05}
M. V. Gurudev Dutt {\it et al.}, \prl{94}{227403}{2005}.

\bibitem{PFreview}
W. T. Pollard, A. K. Felts, and R. A. Friesner, Adv. Chem. Phys. {\bf 93}, 77 (1996).

\bibitem{PF94}
W. T. Pollard and R. A. Friesner, J. Chem. Phys. {\bf 100}, 5054 (1994).

\bibitem{AgarwalLambda}
S. Menon and G. S. Agarwal, arXiv:quant-ph/9902021v1 (1999).

\bibitem{Ou}
B.-Q. Ou, L.-M. Liang, and C.-Z. Li, Opt. Commun. {\bf 281}, 4940 (2008).

\bibitem{TBP}
T. V. Tscherbul and P. Brumer, to be published.

\bibitem{Uzer}
T. Uzer (with an appendix by W. H. Miller), Phys. Rep. {\bf 199}, 73 (1991).

\bibitem{Zaheen}
Z. S. Sadeq and P. Brumer, \jcp{140}{074104}{2014}.

\bibitem{PRA}
T. V. Tscherbul and P. Brumer, \pra{89}{013423}{2014}.




\bibitem{Altenmuller}
T. P. Altenm{\"u}ller, Z. Phys. D {\bf 34}, 157 (1995).

\bibitem{Lloyd}
S. Lloyd and M. Mohseni, New J. Phys. {\bf 12}, 075020 (2012).

\bibitem{PlenioNJP}
A. W. Chin, A. Datta, F. Caruso, S. F. Huelga and M. B. Plenio, New J. Phys. {\bf 12}, 065002 (2010).

\bibitem{FMO}
D. V. Voronine, K. E. Dorfman, B. Cao, and A. Joshi, arXiv:1312.0862v1 (2013).

\bibitem{Leonardo}
J. D. Botero, P. Brumer, and L. A. Pach\'on, in preparation (2014).



\end{thebibliography}
\end{document}